\documentclass[%
reprint,
superscriptaddress,
showpacs,preprintnumbers,
nobibnotes,
amsmath,amssymb,
aps,
prl,
]{revtex4-1}
\usepackage{textcomp}
\usepackage{graphicx}% Include figure files
\usepackage{dcolumn}% Align table columns on decimal point
\usepackage{bm}% bold math
\usepackage{xcolor}
\begin{document}
	
	%\preprint{APS/123-QED}
	
	\title{High temperature quantum Hall effect in finite gapped HgTe quantum wells}% Force line breaks with \\
%	\thanks{A footnote to the article title}%
\author{T.~Khouri}
\email{T.Khouri@science.ru.nl}
\affiliation{High Field Magnet Laboratory (HFML-EMFL), Radboud University, Nijmegen, the Netherlands} 

\author{M.~Bendias}
\affiliation{Physikalisches Institut (EP3), Universit\"at W\"urzburg, Am Hubland, 97074 W\"urzburg,Germany}

\author{P.~Leubner}
\affiliation{Physikalisches Institut (EP3), Universit\"at W\"urzburg, Am Hubland, 97074 W\"urzburg,Germany}

\author{C.~Br\"une}
\affiliation{Physikalisches Institut (EP3), Universit\"at W\"urzburg, Am Hubland, 97074 W\"urzburg,Germany}

\author{H.~Buhmann}
\affiliation{Physikalisches Institut (EP3), Universit\"at W\"urzburg, Am Hubland, 97074 W\"urzburg,Germany}

\author{L.~W.~Molenkamp}
\affiliation{Physikalisches Institut (EP3), Universit\"at W\"urzburg, Am Hubland, 97074 W\"urzburg,Germany}

\author{U.~Zeitler}
\affiliation{High Field Magnet Laboratory (HFML-EMFL), Radboud University, Nijmegen, the Netherlands} 

\author{N.~E.~Hussey}
\affiliation{High Field Magnet Laboratory (HFML-EMFL), Radboud University, Nijmegen, the Netherlands} 
\author{S.~Wiedmann}
\email{S.Wiedmann@science.ru.nl}
\affiliation{High Field Magnet Laboratory (HFML-EMFL), Radboud University, Nijmegen, the Netherlands}

	\date{\today}% It is always \today, today,
	
\begin{abstract}
We report on the observation of the quantum Hall effect at high temperatures in HgTe quantum wells with a finite band gap and a thickness below and above the critical thickness $d_\textnormal{c}$ that separates a conventional semiconductor from a two-dimensional topological insulator. At high carrier concentrations we observe a quantized Hall conductivity up to 60\,K with energy gaps between Landau Levels of the order of 25\,meV, in good agreement with the Landau Level spectrum obtained from $\mathbf{k\cdot p}$-calculations. Using the scaling approach for the plateau-plateau transition at $\nu=2\rightarrow 1$, we find the scaling coefficient $\kappa =0.45 \pm 0.04$ to be consistent with the universality of scaling theory and we do not find signs of increased electron-phonon interaction to alter the scaling even at these elevated temperatures. 
Comparing the high temperature limit of the quantized Hall resistance in HgTe quantum wells with a finite band gap with room temperature experiment in graphene, we find the energy gaps at the break-down of the quantization to exceed the thermal energy by the same order of magnitude.
		
%		\begin{description}
%			\item[Usage]
%			Secondary publications and information retrieval purposes.
%			\item[PACS numbers]
%			May be entered using the \verb+\pacs{#1}+ command.
%			\item[Structure]
%			You may use the \texttt{description} environment to structure your abstract;
%			use the optional argument of the \verb+\item+ command to give the category of each item. 
%		\end{description}
\end{abstract}
	
%	\pacs{Valid PACS appear here}% PACS, the Physics and Astronomy
	\maketitle
	
	%\tableofcontents
	
\section{\label{sec:level1}Introduction}
The quantum Hall effect (QHE) \cite{Klitzing1980a} is a universal phenomenon, which occurs when two-dimensional (2D) metallic systems are subjected to a perpendicular magnetic field. 
The magnetic field splits the constant density of states into discrete Landau Levels (LLs) which are separated by energy gaps $\Delta E$. When the Fermi-energy $E_\textnormal{F}$ is in the gapped regions, the Hall conductance is quantized to integer multiples of $e^2/h$, where $h$ is Plancks constant and $e$ the electron charge. This quantization has been observed in a wide range of semiconductor heterostructures such as GaAs/AlGaAs, Si-MOSFETs and SiGe \cite{Ando1982a,VonKlitzing1986,Tobben1992}. The origin of the QHE can be explained on the basis of localised and extended states that occur in the spectrum of impurity broadened LLs \cite{Laughlin1981,Aoki1981}. In the centre of the LLs, extended states exist which lead to a metallic behaviour while in the vicinity of localised states, the bulk behaviour is insulating. Although bulk states in between LLs are localised, dissipationless 1D edge channels are formed which dominate the transport properties in this regime. The consequence is a quantized Hall resistance accompanied by a vanishing longitudinal resistance. 
In order to observe this quantization, low temperatures ($T\lesssim 10$\,K) are, in general, necessary to prevent thermal occupation of extended states in higher LLs. This limit has recently been overcome by a new class of systems which exhibit a linear dispersion. One of the most famous systems with a linear dispersion is the zero gap semiconductor graphene, where the QHE has been observed at room temperature using a magnetic field of 29\,T \cite{Novoselov2007}.
The observation of a quantized resistance at these high temperatures is based on the peculiar nature of charge carrier quantization in a magnetic field which is given by $E_{N}=\pm v_\textnormal{F}\sqrt{2 e \hbar B N}$, where $v_\textnormal{F}$ is the Fermi-velocity, $B$ is the magnetic field, $\hbar$ the reduced Planck constant and $N$ the Landau Level index. For a typical Fermi-velocity in graphene of $v_\textnormal{F}\approx 10^6$\,m/s$^{-1}$, the energy gap between the lowest LL ($N$=0) and the first excited one ($N$=1) is $\Delta E \approx 195\,\textnormal{meV}$ at 29\,T, exceeding the thermal energy at room temperature ($k_\textnormal{B} T \approx 25$\,meV) by almost one order of magnitude. Unlike in conventional systems, where the second sub-band is relatively close to the first, the second sub-band in graphene is not occupied up to very high temperatures thereby supporting the condition for the observation of the QHE. \\
Similar conditions are present in HgTe quantum wells (QWs). A remarkable property of these so-called type-III QWs is that by tuning the quantum well thickness $d_\textnormal{QW}$ above a critical thickness $d_\textnormal{c}= 6.3$\,nm, a transition from a semiconductor to a topological insulator (TI) is achieved \cite{kanig_quantum_2007}. At $d_\textnormal{QW}=d_\textnormal{c}$, the conduction and valence bands touch each other which leads to a single valley gapless 2D Dirac-fermion system \cite{Buttner2011,Kozlov2015} and QHE can be observed up to nitrogen temperatures \cite{Kozlov2014}. In contrast to conventional semiconductors, HgTe QWs with a finite (bulk) band gap, also have a highly non-parabolic dispersion, approaching a linear energy-momentum relation at finite $k$, and are described by the Dirac Hamiltonian \cite{bernevig_quantum_2006-1}. The presence of a small but finite bulk band gap affects the LL dispersion and consequently the energy gaps in the LL spectrum compared to zero-gap systems \cite{Kozlov2014}.\\

In this paper we investigate the high-temperature QH regime of HgTe quantum wells with a finite band gap with $d_\textnormal{QW}$ below and above $d_\textnormal{c}$. The combination of the dispersion relation, leading to charge carriers that obey the Dirac equation, with a second sub-band that is more than 100\,meV above $E_\textnormal{F}$, makes our system ideal to study the QHE up to high temperatures with only one occupied sub-band.

\section{Sample Characterisation}
\begin{figure}
 \includegraphics[width=\linewidth]{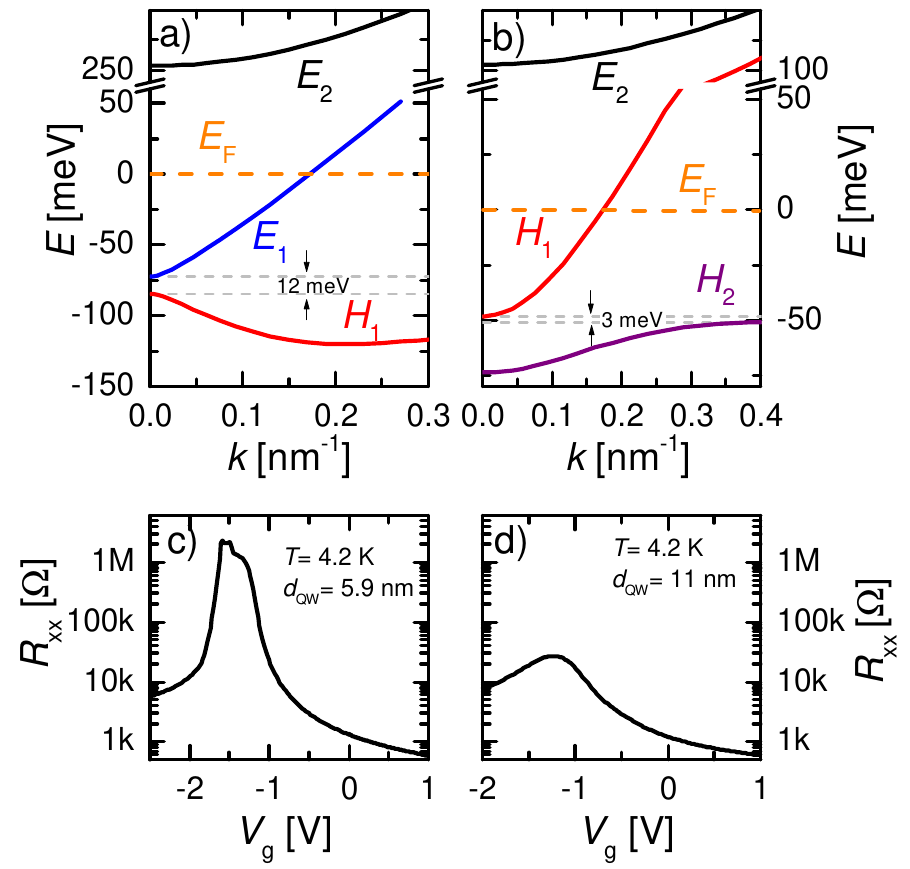}%
 \caption{\label{Fig1}Band structure calculations performed with the $\mathbf{k\cdot p}$-model using a $8\times 8$ Kane-Hamiltonian for a (a) 5.9\,nm thick and (b) 11\,nm thick HgTe QW. Here the lowest two electron like ($E_1$ and $E_2$) and hole like ($H_1$ and $H_2$) sub-band are shown. 
The Fermi-energy at electron densities where QH experiments were performed (see Fig.\,\ref{Fig2}) of (a) $n_\textnormal{s1}=4.59\times 10^{11}$\,cm$^{-2}$ and (b) $n_\textnormal{s2}=4.66\times 10^{11}$\,cm$^{-2}$ is marked by dashed lines. These densities correspond to constant gate voltages of $V_\textnormal{g,1}=$ 1\,V and $V_\textnormal{g,2}=$ 1.4\,V, respectively. Gate dependent measurements at $T=4.2$\,K of the longitudinal resistance $R_\textnormal{xx}$ for the (c) 5.9\,nm thick and (d) 11\,nm thick samples.}
 \end{figure}

Our samples were grown by molecular beam epitaxy (MBE) in the [001] direction and were structured into Hall bars of dimensions $L\times W$ of ($600\times 200$)\,\textmu m$^2$ and ($30\times 10$)\,\textmu m$^2$, respectively. The quantum well thicknesses of the two samples are $d_\textnormal{QW}= 5.9$\,nm (sample~1) and $d_\textnormal{QW}=11$\,nm (sample~2), respectively. The calculated energy dispersions $E(k)$ using a $\mathbf{k\cdot p}$-model with a  $8\times 8$ Kane-Hamiltonian \cite{Novik2005} are shown in Fig.\,\ref{Fig1} (a) and (b) where the first ($E_1$) and second ($E_2$) electron- and  hole-like ($H_1$ and $H_2$) sub-bands are plotted.
Both systems possess a finite bulk band gap and while sample~1 is a trivial semiconductor with a direct band gap of 12\,meV, sample~2 is a 2D TI with an indirect band gap of 3\,meV  and an inverted band ordering giving rise to helical edge states at zero magnetic field \cite{kanig_quantum_2007,brune_spin_2011}. This difference can be seen experimentally when tuning the Fermi-energy $E_\textnormal{F}$ with the top-gate through the band gap while measuring the low-temperature ($T= 4.2$\,K) longitudinal resistance $R_\textnormal{xx}$ (see Fig.\,\ref{Fig1} (c) and (d)).
Compared to the expected insulating behaviour of sample~1 ($R_\textnormal{xx} \geq$ 1\,M$\Omega$), sample~2 has a significantly reduced resistance ($R_\textnormal{xx}\approx 27$\,k$\Omega$) when $E_\textnormal{F}$ is in the bulk band gap. This value is above the expected quantization for a 2D TI due to the formation of charge puddles in larger samples ($L\gtrsim 1$\,\textmu m) which can lead to backscattering \cite{kanig_quantum_2007,vayrynen_helical_2013,Konig2013a} as well as the thermal activation of bulk carriers over the narrow band gap of 3\,meV. \\

To study the QHE at high temperatures we tune the Fermi energy $E_\textnormal{F}$ deep into the conduction band where the dispersion is nearly linear. We achieve this by applying gate voltages of $V_\textnormal{g,1}=$ 1\,V (sample~1) and $V_\textnormal{g,2}=$ 1.4\,V (sample~2) which yields almost equal charge carrier concentrations of $n_\textnormal{s,1}=4.59\times 10^{11}$\,cm$^{-2}$ and, $n_\textnormal{s,2}=4.66\times 10^{11}$\,cm$^{-2}$, respectively. In this regime our samples have mobilities of $\mu_1= 67\,800$ cm$^2$/Vs and $\mu_2=82\,400$ cm$^2$/Vs, as determined from the zero-field resistivity.
For both carrier concentrations $E_\textnormal{F}$ is more than 100\,meV below the second electronic sub-band $E_2$ as can be seen in Fig.\,\ref{Fig1} (a) and (b) from which we exclude thermal excitation of higher sub-bands contributing to transport in our experiment. The nearly linear dispersion in combination with a single occupied sub-band up to high energies are perfect conditions for the observation of QHE at high temperatures.\\
For the experiment we used standard four terminal lock-in techniques and carefully chose our excitation to prevent heating of the samples. We used a $^3$He-system to access a temperature range from 0.3\,K to 80\,K in magnetic fields up to 30\,T. 

\section{Results and discussion}
\begin{figure*}
 \includegraphics[width=\linewidth]{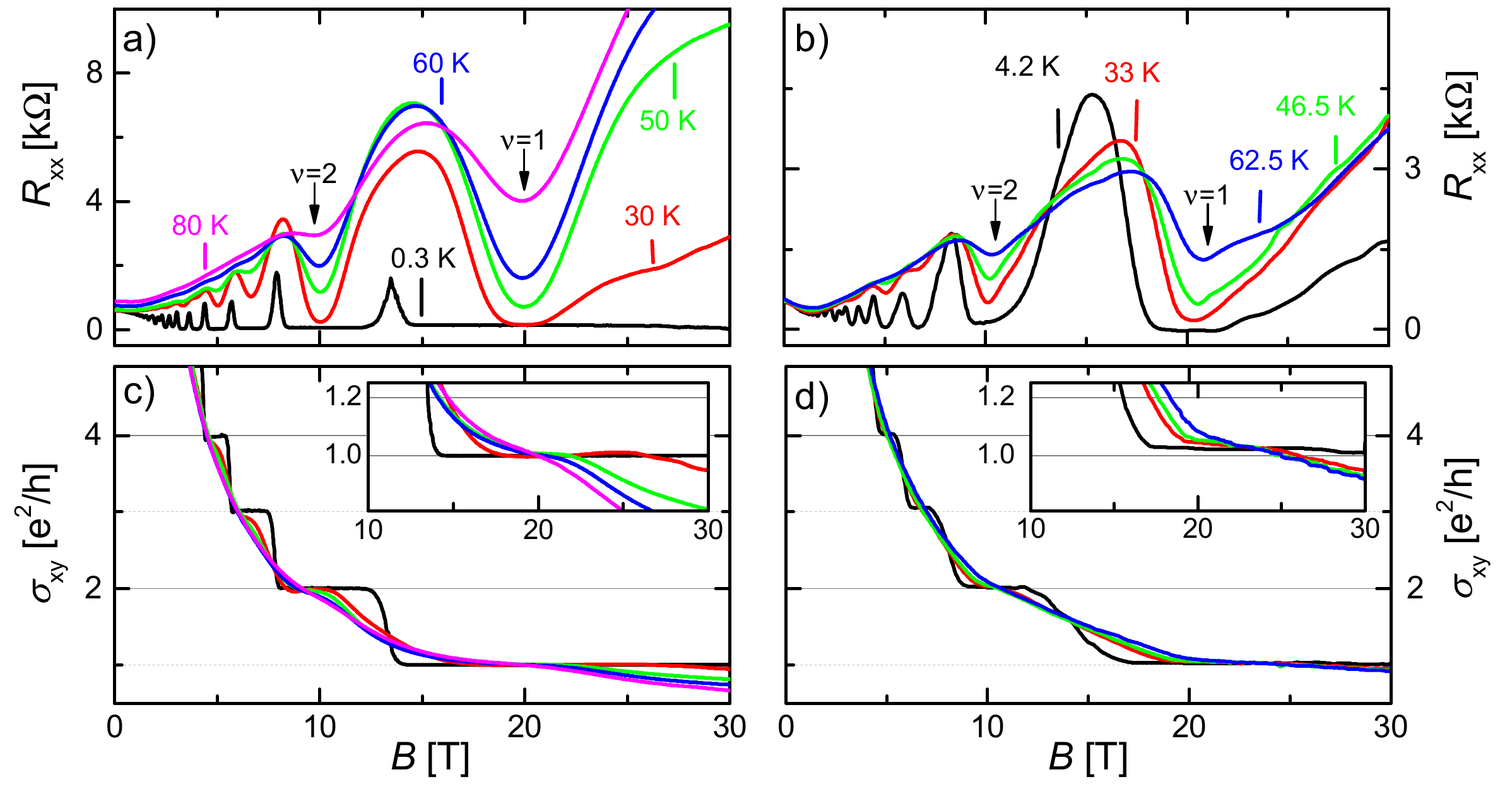}%
 \caption{\label{Fig2}Measurements of the longitudinal resistances $R_\textnormal{xx}$ of the (a) 5.9\,nm thick and (b) 11\,nm thick QWs at constant electron densities $n_\textnormal{s,1}=4.59\times 10^{11}$\,cm$^{-2}$ and $n_\textnormal{s,2}=4.66\times 10^{11}$\,cm$^{-2}$ at different temperatures. The black arrows mark the position in magnetic field of filling factors $\nu$= 1 and $\nu$= 2. In (c) and (d) the corresponding Hall conductivities $\sigma_\textnormal{xy}$ are shown. The insets show a magnification of $\sigma_\textnormal{xy}$ at filling factor 1.}
 \end{figure*}

\begin{figure}
 \includegraphics[width=0.95\linewidth]{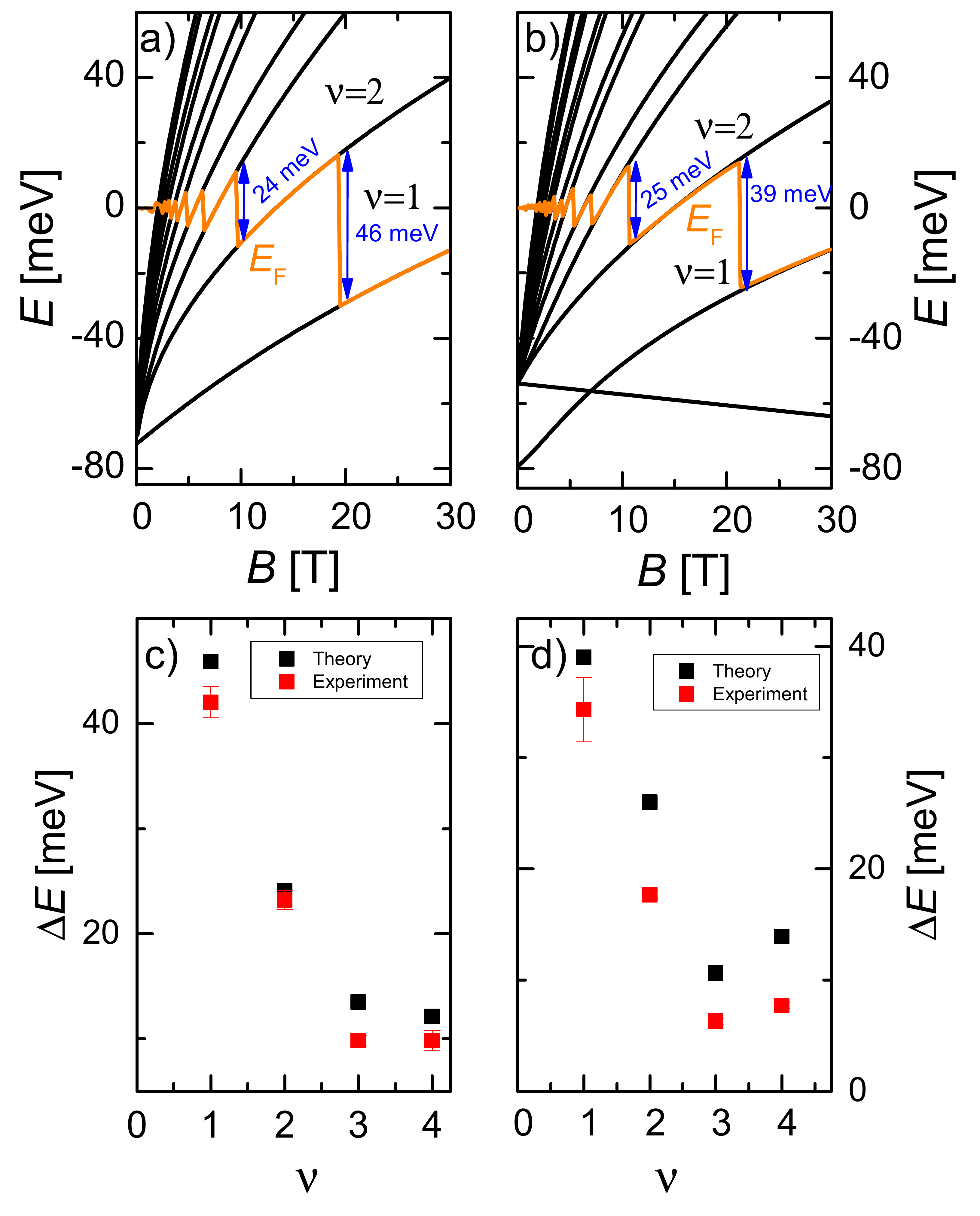}%
 \caption{\label{Fig3} Calculated LL-dispersion for an (a) 5.9\,nm thick and a (b) 11\,nm thick HgTe quantum well. The position of the Fermi energy $E_\textnormal{F}$ is marked by the orange line. In (c) and (d) the corresponding experimentally extracted and theoretical calculated energy gaps are plotted for the lowest four filling factors.}
 \end{figure}

The main results of our measurements are summarized in Fig.\,\ref{Fig2} where we present data which represent the overall behaviour of our samples.
Fig.\,\ref{Fig2} (a) and (b) show measurements of $R_\textnormal{xx}$ of the two samples at constant charge carrier concentrations and different temperatures. From additional measurements of $R_\textnormal{xy}$ and the known geometry of the Hall bars, we determine the corresponding resistivities $\rho_\textnormal{xx}$ and $\rho_\textnormal{xy}$ from which we calculate the Hall conductivities $\sigma_\textnormal{xy}= \rho_\textnormal{xy}/(\rho_\textnormal{xy}^2+\rho_\textnormal{xx}^2)$ plotted in Fig.\,\ref{Fig2} (c) and (d).
The insets show a magnification of the region around filling factor $\nu=1$ where $\nu$ is defined as $\nu= n_\textnormal{s}/n_\textnormal{L}= n_\textnormal{s} h/eB$ where $n_\textnormal{s}$ is the charge carrier concentration and $n_\textnormal{L}$ the degeneracy of the Landau levels. \\
We observe pronounced Shubnikov-de Haas oscillations in the displayed temperature range accompanied by plateaus in $\sigma_\textnormal{xy}$ at $\nu =1$ up to 60\,K and 46.5\,K for samples 1 and 2, respectively.\\
From the temperature dependence of the minima in $R_\textnormal{xx}$, we extract the activation gaps $\Delta E$ between adjacent LLs with a Fermi-Dirac fit and compare the results with theoretical calculations of the Landau Level dispersions shown in Fig.\,\ref{Fig3} (a) and (b). In contrast to the LL fan chart of the 5.9~nm quantum well where all LLs show a positive dispersion, the inverted sample
exhibits a LL crossing of one electron-like and hole-like level at around 8\,T which the hallmark of a 2D inverted system \cite{Buttner2011}.

The experimentally and theoretically obtained energy gaps are in reasonable good agreement as shown in Fig.\,\ref{Fig3} (c) and (d) and the overall behaviour of the sample is well described by our $\mathbf{k}\cdot \mathbf{p}$-model. While the calculations yield energy gaps of $\Delta E_{1,\nu=1}\simeq 46$\,meV at $B\simeq 20$\,T and $\Delta E_{2,\nu=1}\simeq 39$\,meV at $B\simeq 21$\,T for sample~1 and 2, respectively, the extracted activation energies are slightly smaller and determined to be $\Delta E_{1,\nu=1}\simeq (42\pm 1.5)$\,meV and $\Delta E_{2,\nu=1}\simeq (34\pm 2.9)$\,meV. This small difference can mainly be attributed to the simplicity of our calculations where we assume a infinitely small LL width. In reality, scattering from impurities or dopants lead to a broadening of the LLs resulting in smaller energy gaps than our theoretical estimates, as observed. Despite the broadened LLs the energy gap for the lowest filling factor still exceeds the thermal energy at room temperature of $k_\textnormal{B}T\approx 25$\,meV. We furthermore note that the energy gaps are larger than in conventional 2D-systems but are still almost an order of magnitude smaller than in graphene due to a smaller Fermi-velocity ($\simeq 5\times 10^5$\,m/s) \cite{Pakmehr2014,Ludwig2014,Kvon2011} compared to graphene but with a large Zeeman-splitting $\Delta E_{\textnormal{Z}}= g^* \mu_\textnormal{B} B$ of the LLs in HgTe, where $g^*$ is the effective Lande Factor and $\mu_\textnormal{B}$ the Bohr magneton. Therefore, the energy gap for filling factor $\nu=1$ is $\Delta E_{\nu=1}= v_{\textnormal{F}} \sqrt{2\hbar eB}-\Delta E_{\textnormal{Z}}$ and for $\nu=2$ $\Delta E_{\nu=2}= \Delta E_{\textnormal{Z}}$. At the Fermi-Energy, the Landau Level dispersions are very similar and g$^*\simeq$ 20 for both samples. A further increase in charge carrier densities and magnetic field range would, due to the nature of the LL-dispersion, only slightly increase the energy gaps and the maximum $\Delta E$ remains in the order of 50\,meV. Thus, the temperature range where we still observe QHE is largely reduced compared to graphene. Interestingly, the ratio of the energy gap $\Delta E$ to the thermal energy where $\sigma_\textnormal{xy}$ is still quantized is comparable to the ratio of $\Delta E/k_\textnormal{B}T\approx 8$ measured in graphene.\\

Although our system has large energy gaps between adjacent LLs, a necessary but not sufficient condition for the observation of QHE at high temperatures, we need to consider localisation effects of charge carriers. In disordered systems, charge carriers in the tails of the LLs are localised, while extended states exist only at the centre of each LL. Because of these localised states, the Fermi energy moves smoothly through the energy gap and a plateau-like Hall resistance is observed in measurements. A widely used approach to study localisation is to investigate the scaling behaviour of the unique insulator-metal transition which occurs when the energy crosses from localised to delocalised states \cite{Huckestein1995a,Abrahams1979}. Within the finite-size scaling theory \cite{Huckestein1995a}, it is possible to observe scaling behaviour in the temperature dependence of the slope of the plateau transition; specifically the maximum of the derivative of the Hall resistance scales with the temperature as
\begin{equation}
(\textnormal{d}R_\textnormal{xy}/\textnormal{d} B)^\textnormal{max} \propto T^{-\kappa},
\end{equation}
with $\kappa= p/2\gamma$, with $\gamma$ the critical localisation length exponent and $p$ the scattering exponent. A same power law dependence holds for the temperature dependence of the full width half maximum (FWHM) of the $R_\textnormal{xx}$ peaks which we denote as $\Delta B_{R_\textnormal{xx}}$. While the universal scaling theory predicts $\kappa$ and $p$ to be universal it, is still a controversial topic within the literature \cite{Dodoo-Amoo2014,Koch1991a,Hwang1993,Wei1988}.
Since we are able to access a wide temperature range in which we still observe a quantized Hall conductivity it is interesting to compare this scaling behaviour with previous studies.\\

\begin{figure}
 \includegraphics[width=0.8\linewidth]{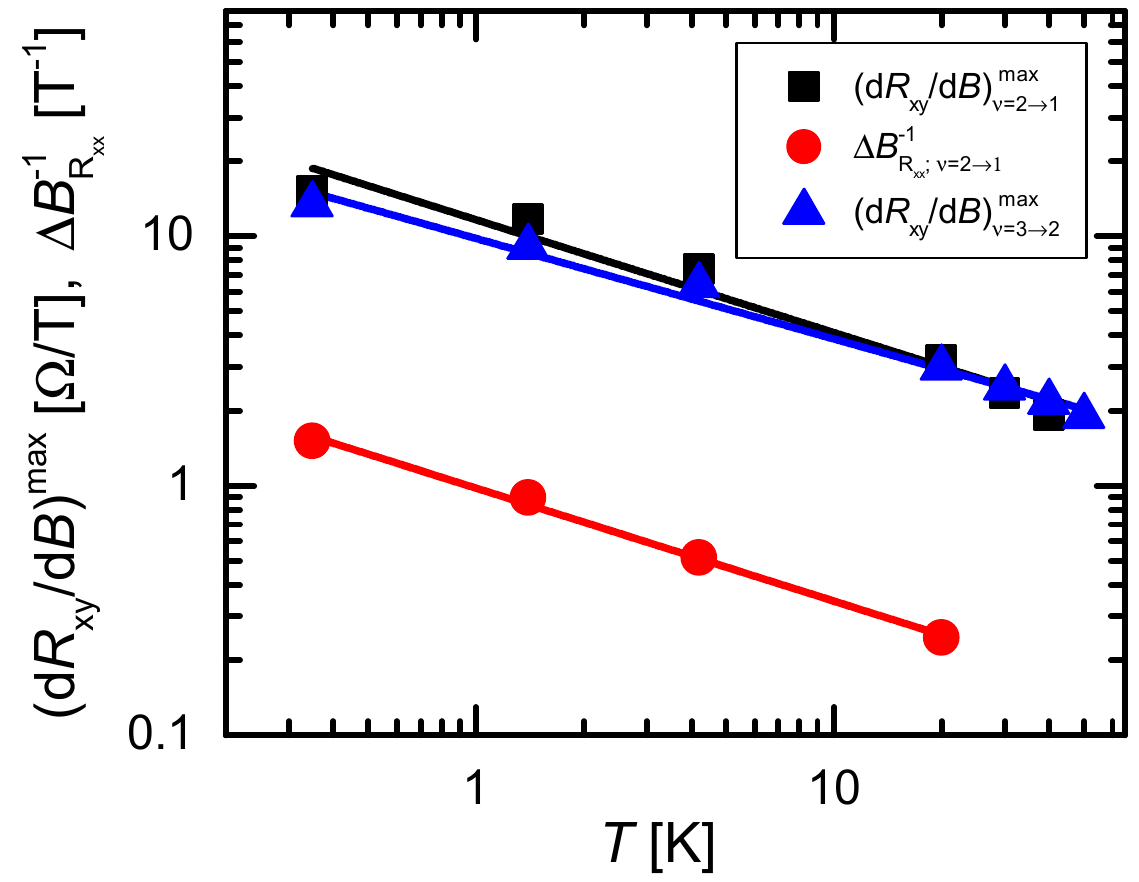}%
 \caption{\label{Fig4} Temperature dependence of the maxima of $(\textnormal{d} R_\textnormal{xy}/\textnormal{d} B)$ and the FWHM of the $R_\textnormal{xx}$ peaks $\Delta B_{R_\textnormal{xx}}$ for the 5.9\,nm thick sample. From the slope of the linear fits $\kappa= p/2\gamma$ can be determined with $(\textnormal{d} R_\textnormal{xy}/\textnormal{d} B)^{\textnormal{max}} \propto \kappa$ and $\Delta B_{R_\textnormal{xx}} \propto -\kappa$.}
 
 \end{figure}

As shown in Fig.\,\ref{Fig4} the analysis of sample~1 of $(\textnormal{d} R_\textnormal{xy}/\textnormal{d} B)^\textnormal{max}$ yields $\kappa=0.45 \pm 0.04$ for the transition from $\nu = 2 \rightarrow 1$ in good agreement with the value $\kappa= 0.45 \pm 0.02$ extracted from  $\Delta B_{R_\textnormal{xx}}$. For the $\nu = 3 \rightarrow 2$ transition $(\textnormal{d} R_\textnormal{xy}/\textnormal{d} B)^\textnormal{max}$ yields $\kappa= 0.40 \pm 0.02$ (unfortunately there are not enough data points available for the analysis of $\Delta B_{R_\textnormal{xx}}$ for this transition as the minima quickly rises above zero). 
Assuming $\gamma$ to be universal, we obtain $p= 2.1 \pm 0.2$.
All our values fit within the theory of universal scaling suggesting that our system is described by short range scattering \cite{Li2009}. Furthermore there is no sign of increased electron-phonon interaction which we expect to be present above 10\,K and would lead to a different scaling behaviour \cite{Lin2002,Brandes1994}.
Our scaling analysis at elevated temperatures is consistent with measurements  on graphene \cite{Giesbers2009} and shows no difference from that obtained on conventional 2D systems.
Similar scaling analysis for sample~2 was not conclusive due to the large error bars in the obtained values of $\kappa$ and $p$. The scaling for a 20.3\,nm wide quantum well has recently been published where the principle feasibility of scaling analysis in HgTe has been addressed \cite{Arapov2015}.

\section{Summary}
In summary we have studied the QHE in HgTe QWs with a finite band gap above and below the critical thickness $d_\textnormal{c}$ up to temperatures of the order of 50\,K.
From temperature dependent magneto-transport measurements we extract energy gaps between LL of the order of 25\,meV. The thermal energy at which the Hall conductance is still quantized is almost factor of 8 higher than the energy gap itself showing striking similarities to graphene. 
We did not find any evidence of increased electron-phonon interaction that would alter scaling behaviour of the QHE between $\nu=1$ and $\nu=2$. From the the observed scaling we determined $\kappa= 0.45 \pm 0.04$ for the non-inverted sample with $d_\textnormal{QW}= 5.9$\,nm  in excellent agreement with the universal scaling theory. An interesting subject for further theoretical and experimental studies is whether the high temperature limit  of the QHE is influenced by a difference in localisation strength and can be related to sample disorder or mobility $\mu$.

\begin{acknowledgments}
This work has been performed at the HFML-RU/FOM
member of the European Magnetic Field Laboratory (EMFL)
and has been supported by EuroMagNET II and is part of the research programme of the Foundation for Fundamental Research on Matter (FOM), which is part of the Netherlands Organisation for Scientific Research (NWO).
The W\"urzburg group would like to acknowledge funding from the German Research Foundation (The Leibniz Program, Sonderforschungsbereich 1170 “Tocotronics” and Schwerpunktprogramm 1666), the EU ERC-AG program (Project 3-TOP) and the Elitenetzwerk Bayern IDK “Topologische Isolatoren”.

\end{acknowledgments}

\bibliography{Literature}

\end{document}